\documentclass{article}
\usepackage{spconf,amsmath,graphicx,hyperref}
\usepackage{enumitem}
\usepackage{booktabs} 

\title{Subjective Evaluation of DNN AND Auditory-Model-Based Hearing-Loss Compensation}

\name{\begin{tabular}[t]{c}
Chuan Wen \textsuperscript{1}, Brent Nissens \textsuperscript{1}, Nele De Poortere \textsuperscript{1}, Morgan Thienpont \textsuperscript{1}, Matthias Inghels \textsuperscript{1}, Attila Fráter \textsuperscript{1}\\
\textit{Guy Torfs \textsuperscript{2}, Sarah Verhulst \textsuperscript{1}}
\end{tabular}\thanks{This work was supported by FWO Machine Hearing 2.0 (216318G) and EIC-Transition EarDiTech (101058278).}}

\address{\textsuperscript{1}Hearing Technology, Department of Information Technology, Ghent University, Ghent, Belgium\\
	\textsuperscript{2} IDLab, Department of Information Technology, Ghent University, Ghent, Belgium}

\begin{document}
	%
	\maketitle
\begin{abstract}
    Outer-hair-cell (OHC) loss is a primary deficit of sensorineural hearing loss (SNHL), impairing cochlear amplification and frequency selectivity and thereby elevating hearing thresholds. Biophysically-inspired DNN-based hearing-aid (HA) algorithms have been proposed to compensate for OHC deficits and have shown clear benefits in objective speech intelligibility and quality metrics (e.g. HASPI, HASQI). However, comprehensive subjective validation of these benefits in human listeners is still missing. In this work, we present a subjective evaluation of a biophysically-inspired HA model targeting OHC deficits. The cochlear module of an auditory model was individualized based on each listener's pure-tone audiogram and integrated into a trainable system, which includes the personalized model and a normal-hearing reference model, and the resulting trained HA was evaluated using a Matrix test comparing intelligibility scores for unprocessed and HA-processed noisy speech. The results revealed a significant benefit of the HA model over the unprocessed condition in the range of +1 to +27\%, providing behavioral confirmation of the efficacy of the HA model. This study closes the gap between objective and perceptual evidence for this new generation of DNN-based HA algorithms, paving the way for their integration into next-generation DNN-accelerated chips for hearables and hearing aids.
\end{abstract}
	\section{Introduction}
	\label{sec:intro}
	
	Sensorineural hearing loss severely degrades speech intelligibility, particularly in adverse acoustic environments. A hallmark deficit of SNHL is OHC damage, which elevates hearing thresholds and  broadens auditory filters, thereby compromising frequency selectivity \cite{moore2007cochlear}. While conventional hearing aids restore audibility through frequency-dependent amplification and wide-dynamic-range compression 
	(WDRC) \cite{keidser2011nal}, they fail to compensate for broadened cochlear tuning, leaving speech-in-noise perception largely degraded \cite{baer1993effects}.
	
	Biophysically-inspired, deep neural network (DNN)-based hearing-aid algorithms have recently been proposed \cite{drakopoulos2023neural,Leer2025,wen2025dconnear,gonzalez2026end}, which learn compensation strategies directly from auditory-model responses rather than relying on predefined, rule-based prescriptions.
	These models are trained within an auditory-model-based framework consisting of two parallel pathways: one simulating the time-domain cochlear response (of different characteristic frequencies) of a normal-hearing (NH) system and the other simulating the response of a hearing-impaired (HI) system. Training minimizes the discrepancy between the NH and HI responses, yielding a numerically optimized compensation strategy without imposing predefined constraints. Such bio-inspired HA models have demonstrated clear benefits in objective intelligibility and quality metrics, such as HASPI \cite{kates2014haspi} and HASQI \cite{kates2014hasqi}, suggesting strong potential for improving hearing-aid compensation performance.
	
	Despite this objective promise, they do not reliably predict behavioral benefits in human listeners. Recent subjective testing by  \cite{Leer2025} provided initial perceptual insight, but was limited by a small cohort ($N = 13$) and high baseline performance in the unprocessed condition ($61\%$ words recognized), potentially introducing ceiling effects that mask true compensation gains. Bridging this gap between objective and perceptual evidence is therefore essential before such compensation strategies can be considered validated for real-world use.
	
	In this work, we present a subjective evaluation of the biophysically-inspired dCoNNear-based HA model targeting OHC deficits \cite{wen2025dconnear}. The model is assessed using a Matrix test with 5-word sentences in noise, comparing intelligibility scores between unprocessed and HA-processed noisy speech. Our results demonstrate significant intelligibility gains over the unprocessed baseline, providing direct behavioral evidence for the efficacy of bio-inspired dCoNNear-based compensation strategies.

	\begin{figure}[tb!]
		\centering
		\includegraphics[width=0.3\textwidth,height=\textheight,keepaspectratio]{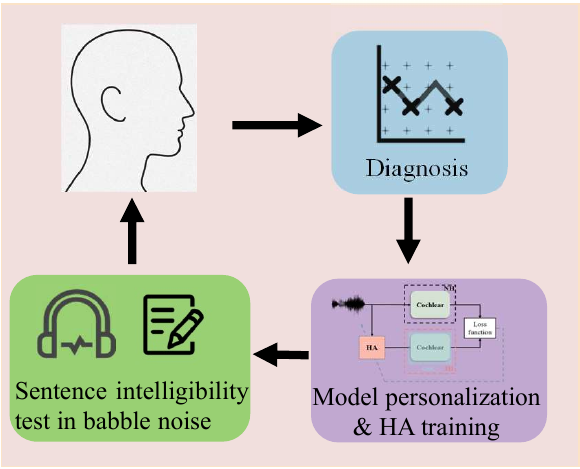}
		\caption{Subjective evaluation procedure for biophysically-inspired hearing aid models:
			(1) Diagnosis using clinical audiogram;
			(2) Model personalization \& HA training;
			(3) Listening tests to measure speech intelligibility in babble noise.}
		\label{fig:scheme}
	\end{figure}
	
	\section{Method}
	\label{sec:method}
	
	The overall experimental workflow comprises three main stages: (i) diagnosis, (ii) model personalization and HA training, and (iii) subjective evaluation, as illustrated in Fig.\ref{fig:scheme}. First, each participant's hearing profile was characterized via pure-tone audiometry. Second, the NH cochlear model was personalized to match the listener's audiometric profile. The resulting individualized HI cochlear model was then embedded into the dCoNNear-based biophysical HA training framework \cite{wen2025dconnear}. Finally, the trained HA models was used to process audio and assess the speech intelligibility benefit behaviorally using a Matrix sentence-in-noise test \cite{luts2014development}.
	
	\subsection{Personalized auditory models}
	Listeners' hearing thresholds were measured across frequencies from 125\,Hz to 16\,kHz and used to individualize the NH dCoNNear cochlear module via transfer learning, yielding a personalized HI cochlear module for each participant.
	
	The cochlear module was based on the dCoNNear architecture \cite{wen2025dconnear}, with a fully convolutional stack of finite-impulse-response (FIR)-like memory blocks. Each memory block integrated a pointwise convolution followed by two distinct dilated depthwise convolutions, which captured historical and future context, respectively. The dCoNNear cochlear module had 12 memory blocks, where each block's history and future modules both used a kernel length of 64 and 256 hidden units, with a hyperbolic tangent (Tanh) activation function.
	
	The reference NH dCoNNear models were trained using simulated responses from the analytical auditory model targets \cite{verhulst2018computational}, following the procedure in \cite{baby2021convolutional}. Cochlear model components were trained on 2,391 utterances drawn from the LibriTTS corpus \cite{zen2019libritts}, with 90\% for training and the remaining 10\% for validation. Personalized cochlear models were then derived from the NH baseline via transfer learning \cite{van2020hearing}, fine-tuned on a smaller subset of only 100 randomly selected LibriTTS utterances, with the mean absolute error (MAE) serving as the training objective.
	
	\subsection{Hearing-aid model training}
	During HA optimization, the parameters of NH and HI cochlear modules within the framework remained frozen. The learning objective was to minimize the mismatch between simulated auditory responses in the NH and HI pathways, ensuring that the HI response to the enhanced signal approximate the NH reference. The HA model shared the dCoNNear architecture, using Tanh activation functions to capture level-dependent compressive amplification, with 12 memory blocks whose history and future modules each used a kernel length of 32 and 256 hidden units.
	
	The HA model was trained on 4,000 speech utterances ($\approx$6\,hours) from the LibriTTS corpus \cite{zen2019libritts} for speaker diversity, split into 3,000 training, 500 validation, and 500 test utterances. All signals were RMS-normalized to 70\,dB SPL and segmented into 1-second frames for training.The loss function was a multi-scale mean absolute error (MAE) loss:
	\begin{equation}
		\mathcal{L}_{\mathrm{total}} = \frac{1}{J} \sum_{j=1}^{J} \frac{1}{M_{j}} \sum_{m=1}^{M_{j}} \mathrm{MAE}\left(\hat{y}_{m}^{(j)}, y_{m}^{(j)}\right)
	\end{equation}
	where $\mathcal{L}_{\mathrm{total}}$ denotes the final multi-scale loss, computed as the mean absolute error (MAE) between the HI response $\hat{y}_{m}^{(j)}$ and the NH response $y_{m}^{(j)}$,
	averaged first across $M_{j}$ segments within each of $J$ distinct temporal scales (10\,ms, 50\,ms, 100\,ms, and 200\,ms), and then across scales.
	
	\subsection{Objective evaluation}
	
	To objectively evaluate the predicted speech-intelligibility and speech-quality benefits of the proposed OHC compensation strategy, we calculated the Hearing-Aid Speech Perception Index version 2 (HASPIv2) \cite{kates2021haspi} and Hearing-Aid Speech Quality Index version 2 (HASQI) \cite{kates2014hasqi}. Both metrics compare an evaluated signal with a corresponding clean reference using an auditory-periphery model that can be parameterized according to the listener's audiogram \cite{kates2022overview}.
	
	The evaluation was performed using a fixed Slope35\_5 audiogram, representing 5~dB HL hearing loss up to 1~kHz, sloping to 35~dB HL at 8~kHz. Stimuli were recorded using a head-and-torso simulator (HATS) positioned at the center of a 1-m-radius ring of eight loudspeakers \cite{manchaiah2024soundscore}. Clean male speech, taken from the LibriVox recording of \textit{The Odyssey} by Homer (Samuel Butler translation),\footnote{\url{https://archive.org/details/odyssey_butler_librivox}} was presented from the frontal loudspeaker at 70~dB SPL, while all eight loudspeakers reproduced multitalker babble, resulting in an overall SNR of $-3$~dB at the HATS position. The same recorded input was used for the unprocessed, OHC, and NAL-NL2 processing conditions.
	
	Both HASPIv2 and HASQI were calculated using the corresponding Slope35\_5 audiogram to parameterize the hearing-impaired auditory model \cite{kates2021haspi,kates2014hasqi,kates2022overview}. HASPIv2 evaluates the similarity of speech-relevant auditory representations between the clean reference and the noisy or processed signal, whereas HASQI quantifies changes related to the speech envelope, temporal fine structure, and long-term spectral characteristics.

	The OHC model was additionally compared with the NAL-NL2 prescription \cite{keidser2011nal}, generated using openMHA v4.18.0 \cite{kayser2022openmha} and version 2023.08 of its NAL-NL2 wrapper. The Slope35\_5 audiogram was mapped onto the default nine-band compressor. To avoid abrupt transitions between adjacent compressor bands, the prescribed gains were interpolated across frequency using monotone cubic Hermite (PCHIP) interpolation on a logarithmic frequency axis.
	
	As summarized in Table~\ref{tab:objective_metrics}, OHC processing substantially increased the HASPIv2 score compared with both the unprocessed and NAL-NL2 conditions. In contrast, the OHC model yielded a lower HASQI score, indicating a trade-off between predicted speech intelligibility and preservation of the signal characteristics captured by the quality metric.
	
	\begin{table}[t]
		\centering
		\begin{tabular}{lcc}
			\toprule
			\textbf{Condition} & \textbf{HASPIv2} & \textbf{HASQI} \\
			\midrule
			Unprocessed      & 0.4581           & 0.2132 \\
			Smoothed NAL-NL2          & 0.5305           & 0.2132 \\
			OHC compensation & 0.9181           & 0.1615 \\
			\bottomrule
		\end{tabular}
		\caption{Objective HASPIv2 and HASQI scores for the evaluated processing conditions at $-3$~dB SNR.}
		\label{tab:objective_metrics}
	\end{table}
	
\begin{figure}[t]
		\centering
		\includegraphics[width=0.7\linewidth]{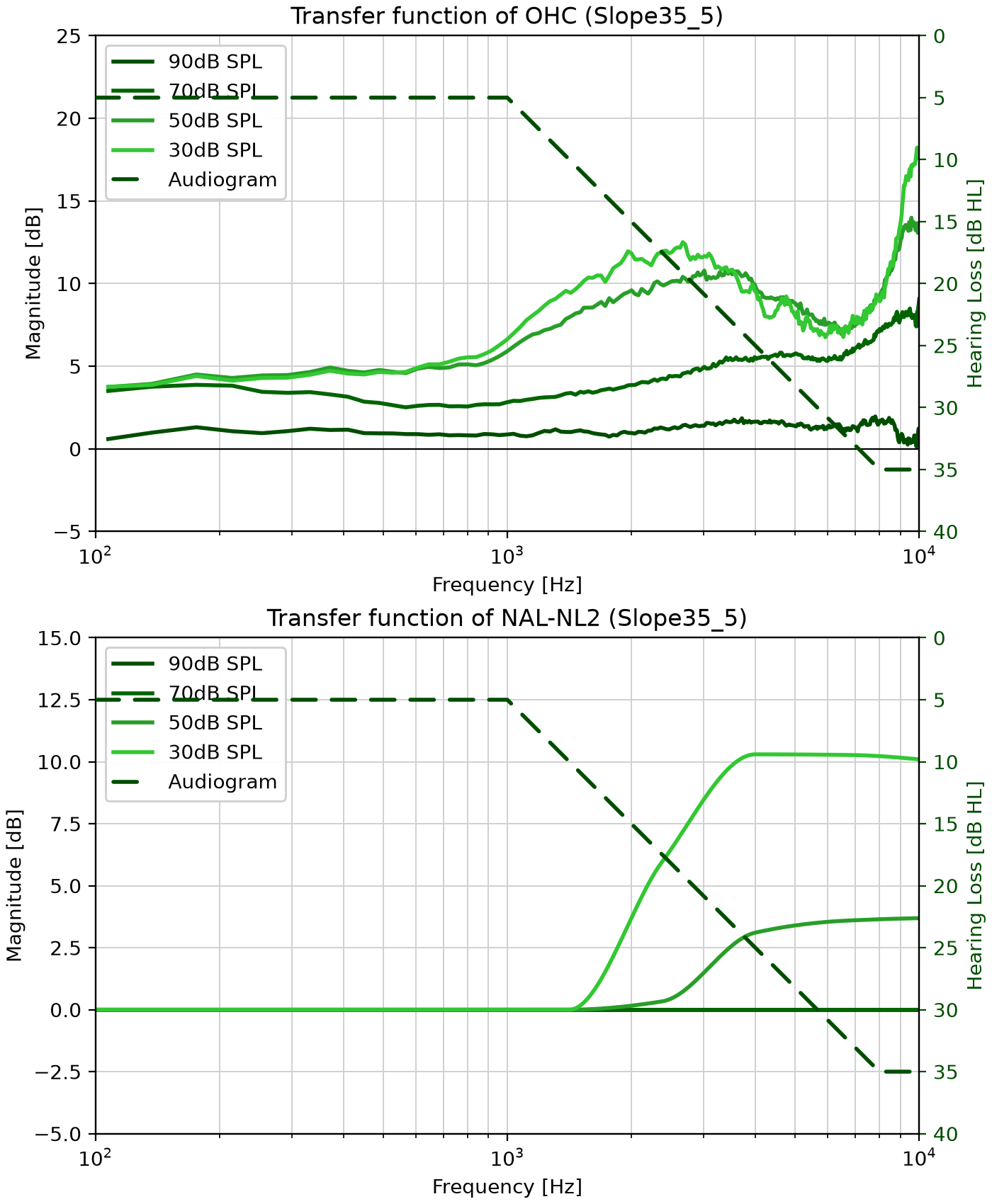}
		\caption{Transfer functions of the OHC model and the smoothed NAL-NL2 algorithm for the Slope35\_5 audiogram.}
		\label{fig:transfer_ohc_nal}
	\end{figure}

	Figure~\ref{fig:transfer_ohc_nal} compares the frequency-dependent transfer characteristics of the OHC model and the smoothed NAL-NL2 prescription for the Slope35\_5 audiogram. Whereas NAL-NL2 derives amplification from an established audiogram-based prescription \cite{keidser2011nal}, the OHC model learns its compensation by minimizing differences between normal-hearing and hearing-impaired auditory-model responses.
	
	\subsection{Subjective evaluation}
	
\begin{figure}[t]
		\centering
		
		\includegraphics[width=0.6\linewidth]{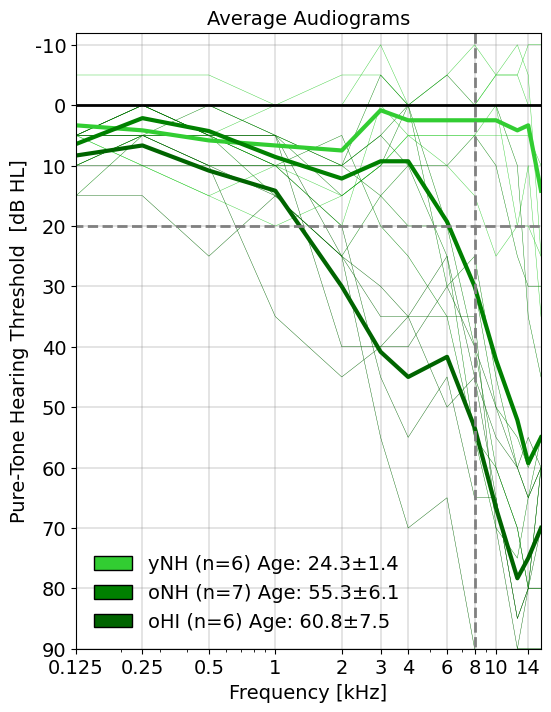}
		\caption{Average audiograms for the three listener groups.}
		\label{fig:audiograms}
		
		\vspace{2mm}
		
		\includegraphics[width=0.6\linewidth]{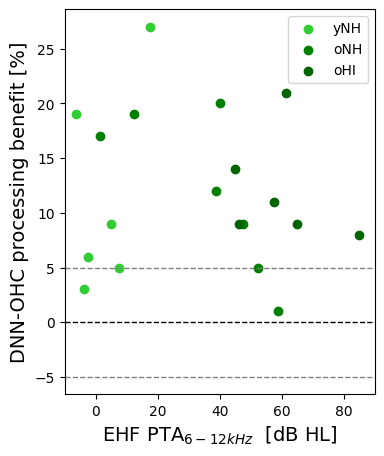}
		\caption{OHC processing benefit (\%) as a function of 
			extended high-frequency pure-tone average (EHF PTA, 6--12 kHz).}
		\label{fig:benefit_vs_ehf}
		
	\end{figure}

	A total of 31 participants were recruited: 9 young normal-hearing (yNH), 14 older normal-hearing (oNH), and 8 older hearing-impaired (oHI) listeners. Individual SNRs were selected to target 40--71\% correct performance in the unprocessed condition and avoid floor or ceiling effects. Only participants whose measured baseline scores fell within this range were included, resulting in 19 participants: 6 yNH, 7 oNH, and 6 oHI listeners. Average audiograms for the three groups are shown in Figure \ref{fig:audiograms}.
	
	All listening tests were carried out in a soundproof booth using Sennheiser HDA300 headphones, with the best ear selected according to each participant’s audiogram. Unprocessed samples were calibrated to 70 dB SPL. To minimize learning effects, each participant underwent a training session consisting of 10 clean and 10 noisy sentences from the Flemish Matrix Test \cite{luts2014development}. All protocols were approved by the UZ Gent medical ethics committee (ON2-2025-0128).
	
	Speech recognition in noise across different processing conditions was calculated as the percentage of words correctly identified. Test stimuli were created by mixing clean Matrix sentences with multitalker babble noise recorded in a restaurant setting \cite{varga1993assessment}.
	Speech and noise were mixed segmentally, using only the active (speech-present) segments of the noise signal together with the RMS level of the corresponding Matrix sentence, following the mixing procedure of the Microsoft DNS Challenge \cite{reddy2021interspeech}. We note that there is an implementation error in the available online SNR mixer of the Microsoft DNS Challenge, causing the SNR-scaling ratio to be applied one time too many and the SNR to have a mean offset of +1dB \cite{reddy2021interspeech}.
	Processed outputs from the DNN-based OHC system are then applied to these speech-noise mixtures, resulting in two distinct test conditions: unprocessed and OHC-compensated.  
	Each condition was assigned to one of the 13 Flemish Matrix lists \cite{luts2014development}, each comprising 20 sentences with five words per sentence. Participants were presented with all possible words in a matrix form (10x5) and selected the words they heard for each sentence.

	\section{Results}
	
	
	To evaluate whether the efficacy of DNN-based OHC compensation depends on OHC damage severity, Fig.~\ref{fig:benefit_vs_ehf} depicts the individual processing benefit, defined as the percentage-point improvement in word recognition relative to the unprocessed baseline, as a function of the extended high-frequency pure-tone average (EHF PTA, 6 to 12~kHz). Elevated audiometric thresholds are generally associated with poorer speech intelligibility, particularly in noisy conditions, making EHF PTA a relevant audiometric marker against which processing benefit can be evaluated.
	
	
	OHC processing yielded a positive intelligibility benefit in nearly all participants, with 17 of 19 listeners (89\%) showing at least a $+5\%$ improvement. Benefits were observed across all three listener groups and across the full range of EHF PTA values, without a clear association between greater hearing loss and larger processing benefit. This may partly reflect the controlled baseline-performance range used to reduce differences in task difficulty. Group-specific benefit ranges are summarized in Table~\ref{tab:subjective_groups}.
	
	\begin{table}[t]
		\centering
		\begin{tabular}{lcc}
			\toprule
			\textbf{Group} & \textbf{$N$} & \textbf{OHC benefit} \\
			\midrule
			yNH & 6 & $+3$ to $+27$ pp \\
			oNH & 7 & $+9$ to $+20$ pp \\
			oHI & 6 & $+1$ to $+21$ pp \\
			\bottomrule
		\end{tabular}
		\caption{Speech-intelligibility benefit of OHC compensation for the three listener groups. Benefits are expressed in percentage points (pp) relative to the unprocessed condition.}
		\label{tab:subjective_groups}
	\end{table}
	
	To confirm this group-level benefit, we conducted a paired $t$-test comparing scores between the Unprocessed and OHC conditions across the 19 participants, appropriate for this fully paired, two-condition design. The improvement was statistically significant ($M_{\text{diff}} = 0.118$, $t(18) = 7.33$, $p < .001$), consistent with the individual-level benefits observed in Fig.~\ref{fig:benefit_vs_ehf}. These results indicate that the biophysically-inspired HA model provided a statistically significant intelligibility benefit for OHC damage, with a large effect size (Cohen's $d_z = 1.68$) confirming that this benefit is robust despite the moderate sample size.
	
	These behavioral findings align with the objective evaluation reported in Section 2.3, where HASPIv2 anticipated a substantially larger benefit for the OHC model than for the unprocessed condition. This convergence between objective and subjective evidence directly addresses the validation gap motivating this study, though the two evaluations were not strictly matched: the objective evaluation used a fixed audiogram and SNR, whereas the subjective test used each
	participant's individual audiogram and SNR. The objective evaluation also benchmarked the OHC model against the NAL-NL2, though the corresponding behavioral comparison remains an important direction for future work. We further note a trade-off in the objective results: the OHC model achieved a lower HASQI score than NAL-NL2, a quality-related dimension not assessed behaviorally here.
	
	\section{Conclusion}
	We presented a subjective evaluation of a biophysically-inspired DNN hearing-aid algorithm targeting OHC deficits, complemented by an objective assessment using HASPIv2 and HASQI. DNN-based OHC compensation provided a consistent, statistically significant intelligibility benefit over the unprocessed baseline, in agreement with the objective predictions obtained in this study. Benefits were observed across the range of OHC-loss severity represented in our sample. Future work should extend the behavioral evaluation to directly compare this approach with current widely used compensation strategies such as NAL-NL2.

\bibliographystyle{IEEEbib}
\bibliography{strings,refs}

\end{document}